\documentclass[12pt]{iopart}

\usepackage{latexsym,epsfig, amsfonts}

\expandafter\let\csname equation*\endcsname\relax
\expandafter\let\csname endequation*\endcsname\relax
\usepackage{amsmath}
\usepackage{bm}

\def\0gr{{\bf 0} }

\begin{document}

\title{Modelling of limitations of bulk heterojunction 
architecture in organic solar cells II: 3d model}

\author{Jacek Wojtkiewicz}
\address{ Faculty of Physics, University of Warsaw, Pasteura 5, 02-093 Warsaw, Poland}
\ead{wjacek@fuw.edu.pl}

\author{Marek Pilch}
\address{ Faculty of Physics, University of Warsaw, Pasteura 5, 02-093 Warsaw, Poland
}
\ead{Marek.Pilch@fuw.edu.pl}


\begin{abstract}

Polymer solar cells are considered as very promising candidates for development of photovoltaics of the future. They are cheap and easy to fabricate. However, up to now, they possess fundamental drawback: low effectiveness. In the most popular BHJ (bulk heterojunction)  architecture the actual long-standing top efficiency is about 12\% (recent achievements about 15\%). One ask the question how fundamental this limitation is, as  certain theoretical considerations suggest that it should be about two times higher. In our paper we analyze the `geometric factor' as one of possible explanation of relatively low efficiency of BHJ architecture. More precisly, we calculate the effective area of the donor-acceptor border in the random mixture of donor and acceptor nanocrystals and further compare it with an ideal 'brush' architecture. In our previous calculation for the two dimensional model, we have found that the maximal value of geometric factor was about 40\%. In the actual three dimensional model, it turned out that both architectures give very close value of the effective area. So the geometric factor seems to be not significant as a factor limiting efficiency.  Implications of this fact  are discussed: we list two other factors (mentioned but not thoroughly discussed in literature) which can be responsible for limitations of efficiency of BHJ architecture. We  estimate their scale, and suggest that these limitations are inevitable, or at least very hard to overcome. We suggest that return to layer architecture could radically improve efficiency limitations -- however, to make breakthrough, materials with large exciton diffusion length have to  be invented.
\end{abstract}

\maketitle

\section{Introduction}


Organic photovoltaics is considered as one of the most perspective investigational trends in entire topic of new types of solar cells design. Main advantages of the organic photovoltaic cells are: low cost, flexibility and small weight. Regrettably, the price we have to pay so far is low efficiency: for a few years the efficiency record  has been fixed on the level of 12\% \cite{nrel}. Only recently, the efficiency has been improved to about 15\% \cite{Joule}. One  can ask about perspectives to improve this efficiency in order to achieve the performance of the best inorganic  single junction  GaAs cells (30\%), or at least  commercial silicon cells (18-27\% \cite{nrel}). This problem has been raised in numerous papers, see for  instance  \cite{JanssenNelson}, \cite{ScharberSariciftci}, \cite{Forrest}.

In order to recognize various aspects of the problem, let us first remind the basic mechanism of the action of solar cell. The conversion of light into electric current in organic cell is a complex, multistage process.  One can recognize the following main stages of it \cite{JanssenNelson}, \cite{Heeger}.

Basic elements of active layer of a cell are: the electron donor and the acceptor. In most cases, the donor is an organic polymer or oligomer. For the second component, i.e. an acceptor,  fullerenes or their chemical derivatives are used in most cases. In the first stage of photovoltaic action, the donor absorbs  photons of solar light. After absorption,  an exciton is formed (i.e. a bound state of excited electron and a hole). It diffuses to the border between donor and acceptor. On the border, the exciton dissociates onto an electron and a hole. The hole remains confined in the donor, whereas the electron moves to the acceptor. In the last stage, the carriers of electric charge diffuse to  electrodes, where they accumulate. As a result, we observe the voltage between  electrodes.

\begin{figure}
\includegraphics{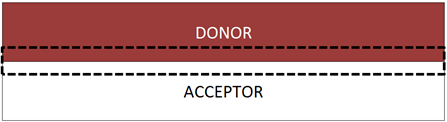}
\caption{\label{Layer2d} Scheme of layer architecture. The characteristic width of layers is 100 nm, and exciton diffusion length is 10 nm. The whole photovoltaic action takes place within region bounded by dashed lines. Excitons formed outside the region bounded by dashed line decay without contribution to the photovoltaic action}
\label{fig:Layer2d}
\end{figure}

An opportunity, which  must be  taken into account in the course of solar cells designing is a short diffusion length of an exciton. In most cases, it is of the order of  10 nanometers, rarely exceeding this value 
\cite{RandRichterCh2}. Historically, the first organic solar cells have been built in a simple layer architecture \cite{Tang} (Fig.~\ref{fig:Layer2d}). The layer thickness should be comparable to the optical penetration depth, so they were of the order of 50-100 nm \cite{Tang}--\cite{Layers3}. It turns out that  solar cells built in the layer architecture exhibit rather limited efficiency (up to 5\%). The most important factor which limits their efficiency is that majority of the excitons decays without dissociation into an electron and a hole before they achieve the donor -- acceptor border.

\begin{figure}
\includegraphics{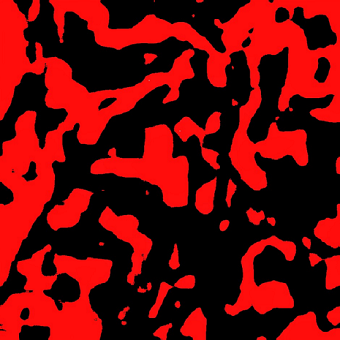}
\caption{\label{BHJa} Very schematic view of the BHJ architecture, frequently encountered in the literature (two-dimensional scheme)}
\label{fig:BHJa}
\end{figure}

Partial solution of this problem is given by the most popular now architecture called BHJ (Bulk HeteroJunction) \cite{BHJ1}-\cite{BHJ5} (Fig.~\ref{fig:BHJa}). In a typical case, the active layer is composed of grains of donor (D) and acceptor (A). The characteristic grain size is of the order of  tens of nanometers. This makes the  area of D-A  contact  to be  large and generated exciton can get the border with an acceptor with high probability. It's great opportunity of the BHJ architecture. The another opportunity is it's simplicity: to prepare an active BHJ blend, it suffices to mix donor's and acceptor's solutions and after evaporation of solvent,  the blend is ready.

\begin{figure}
\includegraphics{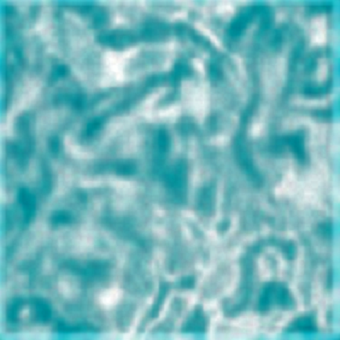}
\caption{\label{BHJb}  Selected view of the BHJ blend structure  supported by TEM images: fuzzy borders between donor and acceptor}
\label{fig:BHJb}
\end{figure}
\begin{figure}
\includegraphics{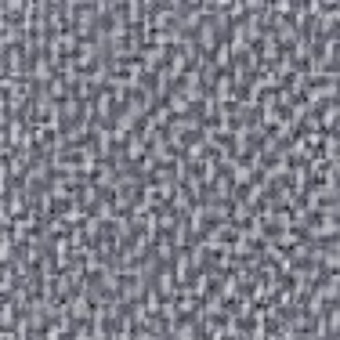}
\caption{\label{BHJc} Selected view of the BHJ blend structure  supported by TEM images: Relatively sharp borders between donor and acceptor}
\label{fig:BHJc}
\end{figure}

However, BHJ architecture has also certain drawbacks. One of them is a creation of  'islands' of donor and acceptor, i.e. an attendance of the grains, which have no connection with any electrode. In such a situation, even if a charge is generated on an 'island', it can't go to the proper electrod. This mean losses in the cell's efficiency. Additional losses are caused by attendance of the 'bad peninsulas', i.e. the donor's grains against the cathode and the acceptor's grains against the anode. It's obvious that there are factors affecting negatively the solar cell's effectiveness in the BHJ architecture. These negative factors were recognized very long ago, but surprisingly  we couldn't find  in the literature  estimates of the scale of these effects.

\begin{figure}
\includegraphics{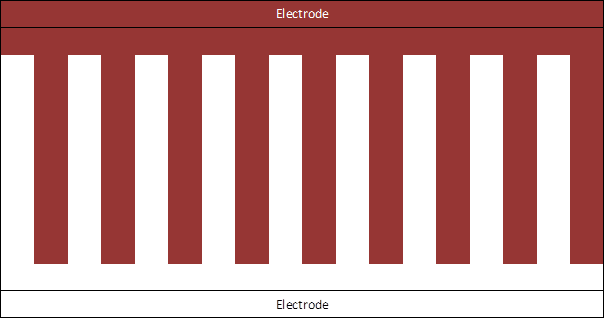}
\caption{\label{comb2d} Scheme of 2d `comb' architecture}
\label{fig:comb}
\end{figure}

\begin{figure}
\includegraphics{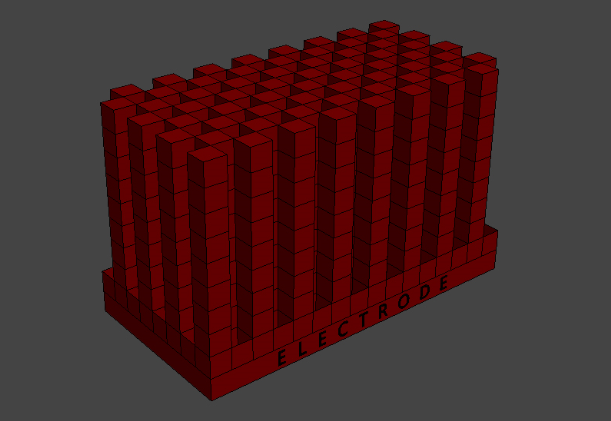}
\caption{\label{BrushHalf} Donor nanocolumns forming the `brush' }
\label{fig:BrushHalf}
\end{figure}

\begin{figure}
\includegraphics{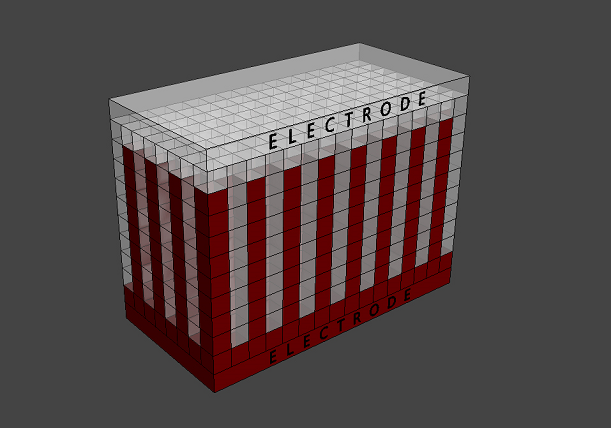}
\caption{\label{BrushFull} Scheme of cell in the `brush' architecture}
\label{fig:BrushFull}
\end{figure}

On the other hand, the architecture which is well-fitted to the exciton's features is so-called `comb' or 'brush' architecture (Figs. \ref{fig:comb},\ref{fig:BrushHalf},\ref{fig:BrushFull}) \cite{Gunes} (the `brush' is a three-dimensional analogon of two-dimensional 'comb'). The area of the contact between donor and acceptor is large, and every generated charge has possibility to achieve proper electrode. The size of the donor/acceptor 'teeth'  should fit the exciton's diffusion length, i.e. their characteristic width should be of the order 10 -- 20 nm. Such devices in principle seem to be available on laboratory scale, but  the fabrication of  well-controlled size brushes in large-scale technology is another matter. Nevertheless, comparison of effective areas of contact between donor and acceptor in the optimal 'brush' architecture versus those of the BHJ architecture seems to be very interesting. 
To express the effectiveness in a more quantitative
manner, we define the `geometric factor' Q, being the quotient of the average
of the area of an active contact between the donor and the acceptor $A_{BHJ}$ and
analogous area in the `brush' architecture B:
\[
Q=\frac{A_{BHJ}}{B}
\]
It is clear that  value of $Q$ has immediate relation with total efficiency of the cell. 

We couldn't find such comparisons  in literature. This opportunity led us to pose the following problem: 

{\em Propose a model -- even crude and oversimplified --   which  could estimate the losses due to presence of the 'islands' and the 'bad pennisulas', and farther to compare geometric factors for cells in the BHJ and the brush architecture.}

In the preceding paper \cite{WP1} we have done first step in such modelling, i.e. we  considered the  two-dimensional situation. In the actual paper, we present three-dimensional model, which is more realistic, but it's more complicated too. It turns out that two- and three-dimensional models lead to different conclusions.

The general setup of the model is as follows. We treat the donor's and acceptor's nanograins as the cubes in a simple cubic lattice. We consider also the second version of the model where  grains are hexagonal prisms occupying cells of three-dimensional hexagonal lattice. In both versions, we assume that the donor's and acceptor's grains are randomly distributed. For every distribution of grains, we have computed area of the iterfaces between the regions occupied by donor and acceptor. We allow  the 'parasitic' effects attendance, i.e. fact that the 'islands' and the 'bad pennisulas' interfaces contribute nothing to production of electricity. This way, so-called 'effective' area of the D-A interface was calculated. In the next step, we compared it with the area of the interface in the brush architecture. And last, we average over random configurations of donor and acceptor grains.

  As a result, it turned out that in the average the 'active' area of the interface in comb architecture was {\em approximately the same} as their counterpart in BHJ. In our opinion, this result has deep implications concerning strategies of development of organic solar cells.

The organization of the paper is as follows. In the Sec. 2 we lay out assumptions of the model and present the simulation's algorithm. In the Sec. 3, we present in some details results obtained. In the Sec. 4, we consider the problem implied by results of our simulations: If areas of the D/A contact in both BHJ and the 'brush' architectures are almost identical, why the total efficiency of BHJ cells are so small?  The Sec. 5 is devoted to summary;  perspectives of further complementary investigations are also sketched.


\section{ The model and the computational algorithm}
\subsection{ Kinds of blends which we simulate}
Fig.~\ref{fig:BHJa} is a  schematic illustration of a BHJ blend, frequently encountered in the literature. Very similar figures can be found in numerous papers, see for instance \cite{ChinNat}, Fig. 1B, where the three-dimensional version is presented. However, one can pose the question: How such schematic figures are related to reality? To settle this question, it is necessary to invoke experimental results. Experimental probing of the BHJ blend structure were performed in numerous papers;  an exhaustive review is \cite{BHJExp}. It turns out that  numerous methods, for instance TEM (Transmission Electron Microscopy) suggest rather different pictures of the structure of BHJ blend. First of all, they are quite diverse. One of situations  encountered is illustrated on Fig.~\ref{fig:BHJb} : Regions occupied by the donor and acceptor form irregular shapes, and interfaces between them are fuzzy. For an example of the TEM image similar to Fig.~\ref{fig:BHJa},  see  Fig. 3D in \cite{ChinNat}.  One encounters also the 'grain' structures, similar to those presented on Fig.~\ref{fig:BHJc}: Regions occupied by the donor and acceptor possess similar size and boundaries between them are relatively sharp. Examples of such structures are presented for instance in  \cite{ChinNat}, Figs. 3A,B,C. 
Our modeling refers to  structures where regions occupied by the donor and acceptor have similar size and shape, and the boundary between them is not fuzzy.  Such structures are encountered mostly in  cases where both donor and acceptor are relatively small molecules (containing up to few hundred atoms). Our model is addressed mainly to blends of such type. On the other hand, polymers -- as a rule -- have very small tendency to crystallization, and our model is less adequate for them.

We make two simplifying assumptions: The donor and   acceptor grains possess the same shape and size (they are cubes -- version 1, or hexagonal prisms -- version 2 of the model).
Moreover, we assume  that interfaces between donor and acceptor domains are sharp. 

\subsection{ Basic technical assumptions}
\begin{figure}
\includegraphics{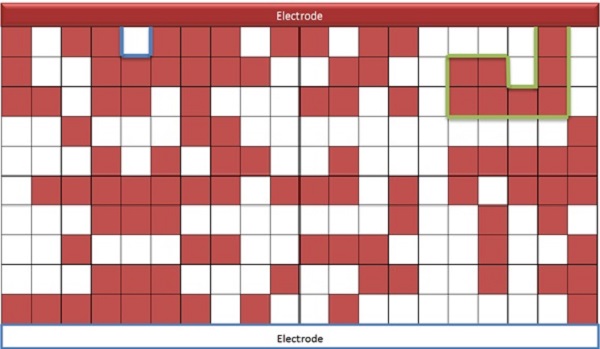}
\caption{\label{BHJsquare2d} The vertical section of the  BHJ layer model. The BHJ blend is formed as a random mixture of cube grains of the donor and acceptor. The blue line indicates the non-active interfaces, whereas the green line concerns the  active interfaces}
\label{fig:BHJsquare2d}
\end{figure}

\begin{enumerate}
\item
The model is three-dimensional one. 
\item
We consider two variants of the model. They are defined on lattices: cubic one (version 1) and three-dimensional hexagonal one (version 2). 
\item
We assume that in version 1 the model of BHJ layer consists of randomly colored white (donor) and red (acceptor) cubes (Fig.~\ref{fig:BHJsquare2d}). Their shape and size correspond to the grain size in some real blends, which is of the order 10 nm (see for instance \cite{RandRichterCh2}, \cite{ChinNat}, \cite{BHJExp}). Of course, one should realize that shapes of grains in real blends can possess very different shapes and sizes. However, there is a quite large group of blends where grains form regular (round or square) shapes of approximately the same size \cite{ChinNat}. Our aim is to simulate such kinds of blend. We took  10 nm as the value of grain size. We assume that every created exciton achieves the donor-acceptor boundary. In version 2, the model of BHJ layer consists of randomly colored hexagon prisms. The diameter of a hexagon corresponds to grain size 10 nm. 
\item
In version 1 we consider subsets of cube lattice with sizes ranging  from 10 x 30 x 30 till 20 x 320 x 320 lattice units. Length of shorter side correspondes to thickness of the BHJ layer, i.e. distance between electrodes. We adopted it as 10, 15 and 20 grains; this way cells with thickness of the BHJ layer of 100-200 nm were simulated. The longer side of the BHJ layer corresponds to characteristic size of an electrode; in a real cell it is widely bigger than thickness of the BHJ layer.  In the version 2 we adopted analogical size of subsets of the hexagon prism lattice. We present sample configurations on section of cubic lattice on Fig.~\ref{fig:BHJsquare2d}  and two-dimensional  illustration of hexagonal lattice on Fig.~\ref{fig:BHJhex2d}.
\item
In cells we can have various ratio of donor and acceptor. We adopted ratios: 1:1, 2:3, 1:2 and 1:4. These ratios have been achieved by the suitable choice of  probabilities for every kind of grain.
\item
The interface between donor and acceptor may be active, i.e. such that the charge created  after excitonÕs decay can get a proper electrode flowing by sequence of connected grains of donor/acceptor. The interface can be also disactive when charges donÕt have such a possibility (i.e. they are trapped in an island without contact with an electrode). Fig. \ref{fig:BHJsquare2d} illustrates examples of active interface (marked by green colour) and example of disactive interface (blue colour) in a blend section.
\item
The area of the active interface is measured and compared with the area of interface in Ôbrush architectureÕ.
\end{enumerate}

\subsection{Computational algorithm}

Now, we present in more details an algorithm devoted to tasks described above. First, we  present an algorithm for cubic lattice in the Subsec.~\ref{subsubsec:2.3.1}. In the next Subsec.~\ref{subsubsec:2.3.2}  we describe an algorithm for three-dimensional hexagonal lattice.  
\subsubsection{Algorithm for the cube lattice}
\label{subsubsec:2.3.1}

We start from declaring a three-dimensional array indexed with non-negative integers
(i, j, k).  In our simulations we used arrays of size ranging from 10 x 30 x 30 till 20 x 320 x 320.  Every individual cell share the two-valued variable, whose interpretation is whether the cell is occupied by the donor or by acceptor. For illustrational purpose, every individual cell of declared array is coloured in one of two colours: white means that a given cell is filled by a grain of donor and red -- by a grain of acceptor.

Parameters of programme are: vertical size $V$ (thickness of the layer); another two sizes $H_x$ and $H_y$; probability $P$ of filling of a given cell by grain of donor. The $P$ value corresponds to ratio of donor and acceptor.

The main steps of the procedure are as follows.
\begin{itemize}
\item[Step 1]: Casting 
 of configurations. For each cell, a random number from an interval [0, 1] is cast. If this number is less than $P$, then a cell is filled with red colour. Otherwise it is filled with white colour. After the step 1, occupations of cells of the lattice by donor and acceptor have been determined. 
 \item[Step 2]: Local connectedness between cells. Examining the charge flow between the sequence of cubes,  touching each other and containing the same content (i.e. the donor or acceptor), we are facing with the following problem. When two touching cubes possess the common wall, it is clear that the charge can flow from one cube to their neighbor. The situation is however different in the case where two touching cubes possess common edge only, or vertex only. In such situations, we have to settle the query: are two touching cubes connected or not?
 
We answer this question in two ways.
\begin{itemize}
\item[Step 2.1]: We assume that the charge flow is possible only in the situation where two cubes possess the common wall. In the other words, the flow of the charge by common edges or vertices are not possible. Within this version of simulation, the algorithm jumps into the step 3.
\item[Step 2.2]: We assume that edges and vertices can contribute to charge flow.
\end{itemize}
 In this case, however, we are facing with the another problem, namely, that flows by connected donor/acceptor cubes are competing ones.  However, such a situation is unstable: By small deformation we obtain the situation where  two cubes containing the acceptor are connected and two other containing the donor are disconnected, or vice versa. (see Fig. \ref{Quest3dCubes}). To settle which one of these two situations happens, we assume that both of them can happen with one-half probability.
In an analogous manner: When two cubes containing one kind of substance touch only by vertex, then they after small perturbation are connected with the probability 0.25, and disconnected with the probability 0.75.
The sketch of implementation procedure is as follows:
\begin{figure}
\includegraphics{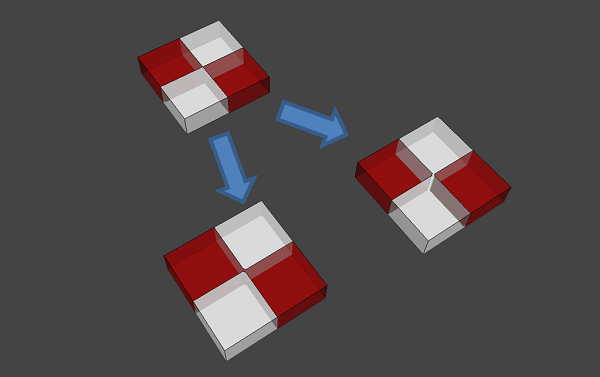}
\caption{\label{Quest3dCubes} In the case where cubes of one type touch by an edge, we randomly choose connection between only 'donor' or only 'acceptor' cubes with equal one-half probability.}
\label{fig:Quest3dCubes}
\end{figure}
  When coloring of cells of the array is finished, the programme again checks all the table to settle the quest of connection between ãgrainsÓ of donor and ãgrainsÓ of acceptor in cases when cells filled by this same colour contact only by edges or only by corners. 
   For each cell programme checks itÕs colour and colours of neighbouring cells. In case when cells filled by this same colour contact only by edge, a random numer from interval [0, 1] is generated. We assumed that in this case cells filled with this same colour are connected with probability 0.5. Obviously, when a pair of identical cells, for example white cells, is connected, then a pair of red cells is not connected. Information about connection or less of connection is scored up. Analogically, in case when cells filled by this same colour are contact only by a corner, quest of connection between them is arbitrated by generating a random number from interval [0, 1] and comparing it with probabilisty 0.25.
\item[Step 3]: When the quest of edge and corner connections between cells is resolved, the programme enters into searching cells filled with red colour, which are connected with the ãredÓ electrode (i.e. with an edge of the BHJ layer which -- as has been assumed -- collects the electrons) by a coherent path composed with the red cells. Analogically are searched the coherent paths composed by the white cells reaching a ãwhiteÓ electrode (i.e. an edge of the BHJ layer which collects holes).
\item[Step 4]:
 In this stage, there are determined these regions from which the charges can flow into an adequate electrode. In step 4 the area of a interface between these regions is calculated. After this  last  step, the total area of the active interface is known and an algorhitm is finishing one simulation.
\end{itemize}
Every four-step simulation is repeated $N$ times (we took $N=100$). After that, standard statistical analysis of active interface area is performed: We calculate the maximum, minimum, average, variance and standard deviation of it. Next, a quotient of the average and the area of interface between donor and acceptor in Ôbrush architectureÕ is computed (in percent).
\subsubsection{Algorithm for hexagon prism lattice}
\label{subsubsec:2.3.2}

\begin{figure}
\includegraphics{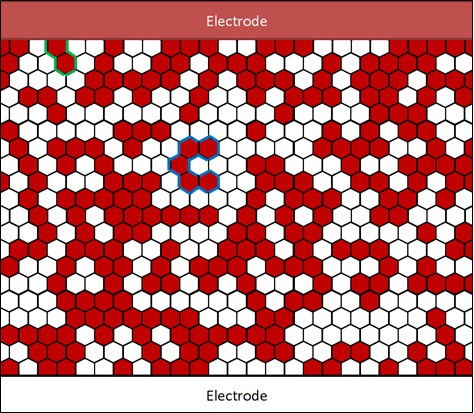}
\caption{\label{BHJhex2d} 
The two-dimensional model of BHJ layer on hexagonal lattice. The BHJ blend is formed as a random mixture of hexagonal grains of the donor and acceptor. The blue line is one of non-active interfaces, whereas the green line is one of active interfaces}
\label{fig:BHJhex2d}
\end{figure}

The algorithm for the hexagon prism lattice  (see Fig. \ref{fig:BHJhex2d} for two-dimensional analogon) is very similar to this for the cube lattice, so we only stress the differences.
\begin{itemize}
\item There are some technical differences with declaration of tables, corresponding to three-dimensional array of hexagon prisms forming a hexagon prism lattice.
\item As in the cube lattice case, we considered two eventualities: one allowing charges movement between cells only by their common wall and the second allowing charges movement by common edges too. In the first eventuality algorhitm just jump to Step 3, skipping Step 2. Because of donor and acceptor grains can touch only by walls and edges and not by vertices, the step 2 is a bit simpler than in cube lattice case -- we must to arbitrate only edges connections quest. 
\end{itemize}


\section{Results}
\subsection{Results for cubic lattice}

We considered lattices of the following sizes: the lengths of shorter edge $V$ were 10, 15 and 20 cells, which corresponded to thickness of the active layer being 100, 150 and 200 nm. The longer edges $H_x$ and $H_y$ were always equal and ranged from 30 to 320 cells. In the Appendix, in Table 1 and Table 2 we present details of results obtained: minimal, maximal and average area of the active interface between donor and acceptor, together with the area for the brush with teeth width equal to 1 x 1 cell. Table 1 concerns the model with `edge' and `vertex' charge flows allowed, and Table 2 present results of the model with forbidden `edge' and `vertex' flows. The values of quotient $Q$ of 'active' interface area and brush area are presented too. The data in the table have been obtained for the proportion 1:1 of the donor and acceptor, and for $N=100$ independent casts of configurations.

\begin{figure}
\includegraphics{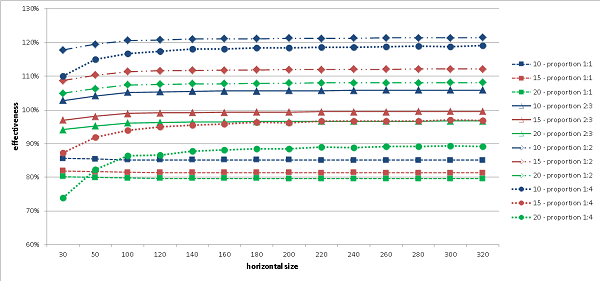}
\caption{\label{czp} Q values for the cube lattice with possibility of charge movement from a cell to another by common edges and vertices. Results for various horizontal and vertical sizes as well as various proportions of donor and acceptor are presented.}
\label{fig:czp}
\end{figure}

\begin{figure}
\includegraphics{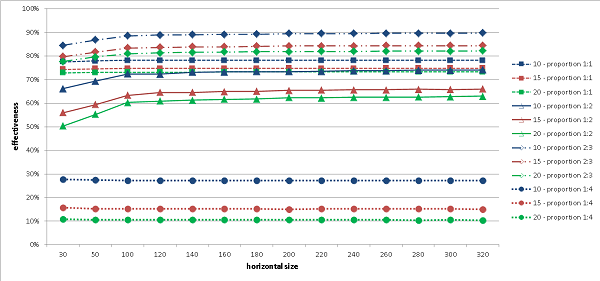}
\caption{\label{cbp} Q values for the cube lattice without charge movement from a cell to another by common edges and vertices. Results for various horizontal and vertical sizes as well as various proportions of donor and acceptor are presented.}
\label{fig:cbp}
\end{figure}

We present the collection of obtained results for the quotient $Q$ in Figs. \ref{fig:czp} and \ref{fig:cbp}.  The Fig. \ref{fig:czp} concern the model with `edge' and `vertex' charge flows allowed and for proportions of the donor and acceptor taking values:  1:1, 2:3, 1:2 and 1:4. We took values in the range of most popular experimental ones. 

For the proportion D/A being 1:1, the largest value of Q was 86\%; for D/A proportion equal to 2:3, the largest value of Q was 106\%; for D/A proportion 1:2, the largest value of Q was 121\% and for D/A proportion 1:4 the largest value of Q was 119\%. High stability of quotient $Q$ with growing horizontal sizes is observed too. The largest values of $Q$ were observed for smallest values of $V$.

The Fig. \ref{fig:czp} concerns the model with `edge' and `vertex' charge flows forbidden. The parameters (proportions of donor and acceptor, and sizes of the system) are the same as for the model with allowed `edge' and `vertex' charge flows. In this case, however, the highest value of $Q$ is smaller than in previous case and equal to 
90\%. 


%

\subsection{Results for hexagonal lattice}

\begin{figure}
\includegraphics{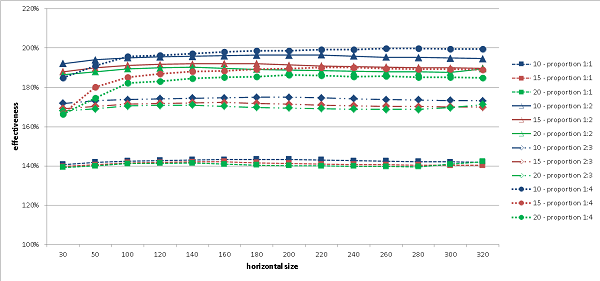}
\caption{\label{hzp}Q values for the hexagonal lattice with possibility of charge movement from a cell to another by common edges. Results for various horizontal and vertical sizes as well as various proportions of donor and acceptor are presented. }
\label{fig:hzp}
\end{figure}

We considered lattices of sizes analogous as previously, i.e. for vertical size (width) being 10, 15 and 20 hexagons and horizontal size ranged from 30 to 320 hexagons. We took four proportions of donor and acceptor: 1:1, 2:3, 1:2 and 1:4. In every case, the averages were calculated from N=100 independent casts, and the quotient Q was computed. We present results on Fig. \ref{fig:hzp}. 

Here $Q$ values are much higher than in cube lattice case: the $Q(H)$ functions tend to certain limit value (depending of D/A proportion and vertical size) which ranges from 140\% to 200\%. Again, quotient $Q$ appeared smallest for proportion 1:1. For instance, for proportion 1:1 and width 10, the maximal value of $Q$  (for the model with `edge' flows allowed) was 143\% (86\% for cubes); for proportion 2:3 and width 10, the maximal value of Q was 175\% (106\% for cubes); for proportion 1:2 and width 10, maximal value of Q was  196\% (121\% for cubes) and for
proportion 1:4 and width 10, maximal value of Q was  200\% (119\% for cubes). For other widths we observe similar interrelations.

\begin{figure}
\includegraphics{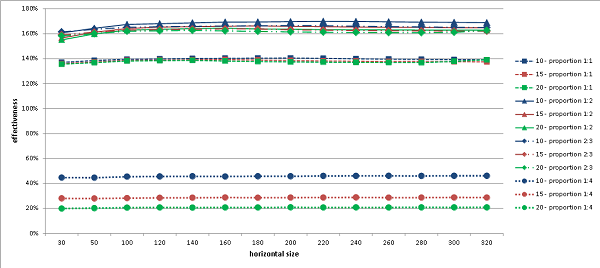}
\caption{\label{hbp}Q values for the hexagonal lattice without possibility of charge movement from a cell to another by common edges. Results for various horizontal and vertical sizes as well as various proportions of donor and acceptor are presented. }
\label{fig:hbp}
\end{figure}

We present also results (Fig. \ref{fig:hbp}) for the model without possibility of charge movement by common edges. As one can expect, the $Q$ value is smaller than for the model with possibility of edge flows. However, results in these two cases do differ not in a drastic manner. 
 

%
 
\subsection{Summary of simulations}

The ultimate goal of our simulations was to find an answer to the question: Which is the area of the interface between Donor and Acceptor in the BHJ architecture  $A_{BHJ}$, compared with the area $B$ of the interface in the 'brushÕ architecture. The answer we have obtained is that the interface area in the BHJ architecture is {\em approximately the same or even bigger} than in the 'brush' architecture. In a more quantitative manner, the quotient $Q =  A_{BHJ}/B$  took the value in the range between  0.74 and 2.0. In our model it turned out that the less efficient proportion of the Donor and Acceptor was 1:1. The value of Q depend of the thickness of the active layer, proportion of the Donor and Acceptor and the shape of their grains.

  An immediate consequence of this opportunity is that the efficiency of the photovoltaic device in the BHJ architecture should be similar or even higher than in the 'brushÕ architecture.

In the most casted configurations, we observed linking of two electrodes by the one kind of component (i. e. by a donor, or by an acceptor). Such a linking is present on exemplary casted configuration on Fig.~\ref{fig:BHJsquare2d}. Presence of such linkings means that created electrons or holes can flow to both electrodes. Of course, it is unwanted process. The remedy to avoid it is well known -- it is to use electron/hole blockades. Our results can be viewed as independent indication of necessity of using electron/hole blockades. Unfortunately, the presence of blockades enhances the internal resistivity, makes the construction of solar cells more complicated and influences the cell efficiency in a negative manner.

{\em Remark.}
In principle, one could consider the statistics of such configurations (where electrodes are linked by one of species forming the active layer) by noticing that this phenomenon is certain kind of {\em percolation}. 
There are many kinds of this phenomenon. Among large amount of the literature devoted to percolation problems, see for instance the textbook \cite{Percolation} for a basic
background in the subject. However, we couldn't find in the literature the specific information needed by
us (i.e. the concrete values of the probability of the ``cube" or ``hexagonal prism'' percolation in finite systems) and this motivated us to undertake our own
simulations.

We observed dependence of the quotient $Q$ of the shape of grains: For cubes, the value of $Q$ was systematically lower than the value for hexagons. It would be interesting to simulate systems with another shapes of grains (by looking at SEM results), consider mixtures of grains with different sizes, or -- more generally (and difficult) -- allow grain shapes not being identical.

\section{Some implications and corollaries}

	Let us look at the obtained results in the context of general state-of-the-art for cells, designed  and constructed in the BHJ architecture.
	
	 At present, efficiencies of BHJ cells does not exceed  12\% and didn't improve during last 6 years \cite{nrel}. It is so despite huge efforts towards enhancement the efficiency. Hundreds of species have been examined (mainly donors). Various aspects of cells have been analysed: chemical, electrochemical, optical...  Thousands papers have been devoted to examine processes which take place within cells,  and to model them. Despite these efforts, the efficiency of BHJ organic cells is still limited by about 12\%. This opportunity leads to conclusion that either there is some fundamental limitation for efficiency of BHJ organic cells, or some crucial factor(s) is (are) skipped in a systematical manner.

On the other hand, there are theoretical indications that the efficiency of organic cells -- even in one-junction simplest version --  can achieve 20-24\% \cite{JanssenNelson}.    These theoretical predictions are based on the Shockley-Queisser estimations \cite{SQ} as well as general optical properties of  donors and acceptors. It seems that there is no basic theoretical obstacles against possibility to enhance the efficiency to be much higher than present top value.
 
For an application of the Shockley-Queisser results, the 'brush' architecture is close to an ideal. The area of the Donor-Acceptor contact achieves the maximal needed value. Every created  electron and  hole has possibility to flow to the proper electrode. Moreover, the process of dissociation of the exciton goes also with very good efficiency \cite{Heeger}. 

The results obtained in our modelling, in turn,  suggest that the area of D/A contact, as well as  efficiency in charge collection take very close values in both architectures: 'brush' and BHJ ones. So -- if only these two factors are important -- there is only marginal difference between  'brush' and BHJ architectures.

It seems that there exist another factors responsible for limitation of efficiency of BHJ cells.

Authors want to pay attention onto two factors, mentioned in the literature but not thorougly discussed, and which seems us to be very important. They are:
\begin{enumerate}
\item the possibility that the D/A boundary is not sharp;
\item presence of the charge surface recombination.
\end{enumerate}

Let us discuss them in more details:
\begin{enumerate}
\item  It is generally assumed that the D/A boundary  is sharp, i.e.  on the one side of the junction there is pure donor and on the other side there is pure acceptor. However, it seems to be an idealization. Real blends are produced by an evaporation of the solvent from the solution containing both the donor and the acceptor. After evaporation, a mixture of nanocrystals of donor and acceptor forms. In such a mixture, the assumption of the sharpness of  the D/A boundary is at least disputable: The faster evaporation is, the more fuzzy the boundary between the grains of
the donor and the acceptor forms. On the other hand, the evaporation cannot
be too slow, as the donor and the acceptor crystals would be too large. So, it seems reasonable to assume that the D/A boundary is not sharp but fuzzy. (Fig. \ref{fig:FuzzyBoundary}). 
\begin{figure}
\includegraphics{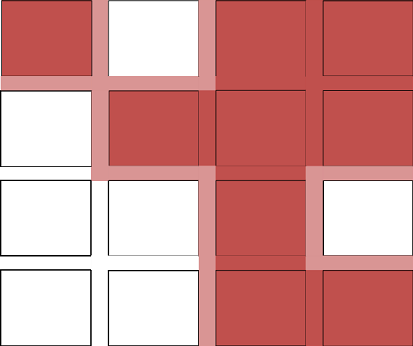}
\caption{\label{FuzzyBoundary} Fuzzy boundary between grains of donor and acceptor.}
\label{fig:FuzzyBoundary}
\end{figure}%
Let us call as 'fuzzy boundary' the region where the concentrations of donor and acceptor are approximately the same. 
The presence of the the fuzzy boundary can influence the cell efficiency in a negative manner in (at least) two ways. The first of them is the 'material losses'. The fuzzy boundary do not participate in the exciton production. How large are losses due the presence of the 'fuzzy boundary'? Consider the cubic lattice and the `brush' architecture.

 If the thickness of the fuzzy region is 5\% of a cells edge, then an amount of pure D and A species decreases to approximately $0.95^3\approx 0.85$, i.e. 85\% of the blend is active in exciton production. If the thickness is 10\%, only 73\%  of the blend is active. We are not aware of papers where detailed examination of the influence of the 'fuzzy boundary' have be investigated, so we pose it as an open problem: 

{\em Examine the presence and structure of the fuzzy boundary, and its influence into the efficiency of the BHJ solar cells}. 
\item The second negative effect is the surface charge recombination. In the BHJ architecture, we have to do with two currents: One is a current of holes flowing within donor and the second -- the current of electrons flowing within the acceptor. The area of contact between donor and acceptor is very large. To be specific, if the characteristic size of grains in BHJ blend is 10 nm, then the area of D/A contact in 1 cm${}^3$ of the blend is of the order of 100 m${}^2$. So it is natural to expect that some of charges will recombine on the D/A boundary, causing the losses of the current. The problem of surface charge recombination has been noticed (see for instance \cite{RoyBose}) but we again are not aware more profound and quantitative discussion of the scale of this phenomenon and its influence to the efficiency of BHJ cells. So we pose the second problem:

{\em Estimate the scale of surface charge recombination in the BHJ architecture.}
\end{enumerate}


\section{Summary and conclusions}

In the paper, we have examined the 'geometric' factor which can influence the efficiency of photovoltaic cells built in the BHJ architecture. More precisely, we have calculated the effective area of the donor-acceptor interface in the random mixture of donor and acceptor nanocrystals and compared it with an ideal 'brush' architecture. As a result, it turned out that the areas in these two kinds architectures are approximately the same. In the other words, influence of the geometrical factor is of small importance.

We have also looked for another factors which make the efficiency of organic cells in BHJ architecture is far less than predicted by certain theoretical considerations, based mainly on the Shockley-Queisser model.
 We made simple estimations suggesting that lower efficiency can be caused by the presence of the regions occupied by the mixture of donor and acceptor (`fuzzy boundary'), where both species are mixed in a molecular scale. We also propose to pay attention to the problem of surface charge recombination. Deeper understanding of  these two effects, and examination of its scale,  should shed light into the problem of (relatively) low efficiencies of cells built in the BHJ architecture.

We do not claim that both factors above are fundamental ones and not possible to overcome. Even if it is true that the fuzzy boundary is massively present in actual BHJ blends, it is possible that one can avoid it by inventing species cristallizing in sharp boundary grains. Also, we don't know effect of surface recombination -- perhaps it is negligible. So we do not claim that BHJ  architecture is futureless.

But one can consider also another directions of development.
One such a possibility, aimed to enlarge the efficiency of photovoltaic devices could be to return to layer architecture. Devices constructed in a layer architecture exhibit lower (however not drastically) efficiency compared with those in BHJ architecture \cite{Tang}-\cite{Layers3}, \cite{RandRichterCh2}. To improve efficiency of  'layer' devices,  one has to solve the main problem: {Find the substance(s), where the exciton diffusion length is comparable to the optical penetration  length}. In more quantitative manner, typical value of optical penetration length is of the order 100 nm \cite{RandRichterCh2}, so one should find the substance where the exciton diffusion length is of the order of 100 nm. It is very difficult task, as in the most of donors or  acceptors used in photovoltaic devices the exciton diffusion length is of the order 10 nm \cite{RandRichterCh2}. But it seems that it is not hopeless, as there are known certain compounds, where    the exciton diffusion length is about 100 nm! \cite{RandRichterCh2}, \cite{Antracen1}, \cite{Antracen2}, \cite{NPD}. Unfortunately, their absorption in the visible range is very small. To find the compound(s) absorbing the light in visible range and possessing the large exciton diffusion length is a great challenge for material research. We are aware of one such a class of compounds possessing  exciton diffusion length exceeding 100 nm -- they are {\em nanotubes}. It would be very interesting to examine their optical properties concerning possibility of their application as active layers in organic solar cells.


\section*{Appendix: Tables: samples of detailed results}

\begin{tiny}

\noindent
Table 1. 
An example of detailed results of simulations for cuboid subsets of cube lattice with possibility of charge movement from a cell to another by common edges and vertices; $V=20$ is vertical size (width), and $H_x = H_y$ are horizontal sizes. The averages are taken from $N=100$ casts. Proportion of the donor and acceptor is 1:1.
 
\noindent
\begin{tabular}{||c||c|c|c|c|c|c|c|c||}
\hline\hline
$V\times H_x \times H_y$&
20x30x30&
20x50x50&
20x100x100&
20x120x120&
20x160x160&
20x220x220&
20x280x280&
20x320x320
\\
\hline\hline
Min. bd. area&
25 639&
71 961&
290 993&
420 152&
748 141&
1 417 462&
2 300 780&
2 642 170
\\
\hline
Max. bd. area&
26 138&
73 044&
292 752&
422 228&
751 159&
1 421 506&
2 297 368&
2 638 844
\\
\hline
Av. bd. area $A_{BHJ}$&
25 839&
72 511&
292 140&
421 218&
750 089&
1 419 770&
2 303 784&
2 643 668
\\
\hline
Variance&
10 588&
42 801&
116 901&
191 018&
467 563&
769 480&
2 132 552&
1 349 229
\\
\hline
brush area $B$&
32 220&
90 700&
366 400&
528 480&
941 440&
1 782 880&
2 890 720&
3 777 280
\\
\hline
$Q = A_{BHJ}/B$ [\%]&
80.2&
79.9&
79.7&
79.7&
79.7&
79.6&
79.6&
79.6
\\
\hline
Std. dev.&
102.9&
206.9&
341.9&
437.1&
683.8&
877.2&
1 460.3&
1 161.6
\\
\hline\hline
\end{tabular}

\vskip1cm

Table 2.  An example of detailed results of simulations for cuboid subsets of cube lattice without charge movement from a cell to another by common edges and vertices; V=20 is vertical size (width), and$ H_x = H_y$ are horizontal sizes. The averages are taken from $ N=100$ casts. Proportion of the donor and acceptor is 1:1.

\noindent
\begin{tabular}{||c||c|c|c|c|c|c|c|c||}
\hline\hline
$V\times H_x \times H_y$&
20x30x30&
20x50x50&
20x100x100&
20x120x120&
20x160x160&
20x220x220&
20x280x280&
20x320x320
\\
\hline\hline
Min. bd. area&
23 121&
65 185&
266 974&
385 423&
687 168&
1 303 277&
2 112 520&
2 760 410
\\
\hline
Max. bd. area&
23 955&
66 789&
269 096&
388 090&
690 662&
1 308 559&
2 119 177&
2 768 171
\\
\hline
Av. bd. area $A_{BHJ}$&
23 476&
66 161&
268 015&
386 683&
689 100&
1 305 538&
2 116 929&
2 765 451
\\
\hline
Variance&
25 141&
80 147&
237 699&
310 108&
689 670&
778 319&
2 650 718&
2 964 954
\\
\hline
brush area $B$&
32 220&
90 700&
366 400&
528 480&
941 440&
1 782 880&
2 890 720&
3 777 280
\\
\hline
$Q = A_{BHJ}/B$ [\%]&
72.9&
72.9&
73.1&
73.2&
73.2&
73.2&
73.2&
73.2
\\
\hline
Std. dev. &
158.56&
283.10&
487.54&
556.87&
830.46&
882.22&
1 628.10&
1 721.90
\\
\hline\hline
\end{tabular}
\vskip1cm 

Table 3. An example of detailed results of simulations for the hexagonal lattice with possibility of charge movement from a cell to another by common edges; V=20 is vertical size (width), and $H_x = H_y$ are horizontal sizes. The averages are taken from $N=100$ casts. Proportion of the donor and acceptor is 1:1.

\noindent
\begin{tabular}{||c||c|c|c|c|c|c|c|c||}
\hline\hline
$V\times H_x \times H_y$&
20x30x30&
20x50x50&
20x100x100&
20x120x120&
20x160x160&
20x220x220&
20x280x280&
20x320x320
\\
\hline\hline
Min. bd. area&
47 898&
134 309&
543 273&
784 151&
1 389 810&
2 612 440&
4 219 257&
5 613 257
\\
\hline
Max. bd. area&
48 851&
136 080&
546 311&
788 510&
1 395 298&
2 618 603&
4 229 019&
5 623 926
\\
\hline
Av. bd. area $A_{BHJ}$&
48 394&
135 194&
544 868&
786 086&
1 392 780&
2 615 160&
4 225 040&
5 619 500
\\
\hline
Variance&
30 330&
119 599&
368 884&
778 422&
1 078 789&
2 345 552&
4 970 218&
5 712 495
\\
\hline
brush area B&
34 738&
96 495&
385 981&
555 812&
988 111&
1 868 150&
3 026 090&
3 952 440
\\
\hline
$Q = A_{BHJ}/B$ [\%]&
139&
140&
141&
141&
141&
140&
140&
142
\\
\hline
Std. dev.&
174&
346&
607&
882&
1 039&
1 532&
2 229&
2 390
\\
\hline\hline
\end{tabular}

\vskip1cm

\noindent
Table 4. An example of detailed results of simulations for the hexagonal lattice without possibility of charge movement from a cell to another by common edges; V=20 is vertical size (width), and Hx = Hy are horizontal sizes. The averages are taken from N=100 casts. Proportion of the donor and acceptor is 1:1.

\noindent
\begin{tabular}{||c||c|c|c|c|c|c|c|c||}
\hline\hline
$V\times H_x \times H_y$&
20x30x30&
20x50x50&
20x100x100&
20x120x120&
20x160x160&
20x220x220&
20x280x280&
20x320x320
\\
\hline\hline
Min. bd. area&
46 547&
130 838&
531 701&
766 415&
1 360 461&
2 558 264&
4 131 746&
5 490 150\\
\hline
Max. bd. area&
47 743&
132 864&
535 368&
771 987&
1 367 090&
2 565 986&
4 143 754&
5 505 006\\
\hline
Av. bd. area $A_{BHJ}$&
47 092&
132 006&
533 161&
769 198&
1 364 400&
2 562 860&
4 139 990&
5 499 220\\
\hline
Variance&
57 010&
136 050&
498 400&
1 076 080&
1 439 917&
2 670 117&
8 980 372&
8 598 509\\
\hline
brush area B&
34738&
96495&
385 981&
555 812&
988 111&
1 868 150&
3 026 090&
3 952 440\\
\hline
$Q = A_{BHJ}/B$ [\%]&
135.6&
136.8&
138.1&
138.4&
138.6&
138.1&
137.7&
137.4\\
\hline
Std. dev.&
238.77&
368.85&
705.97&
1 037.34&
1 199.97&
1 634.05&
2 996.73&
2 932.32\\
\hline\hline
\end{tabular}

\noindent
\end{tiny}

\ack
We want to express our warm gratitude for Prof. Maria Kami\'{n}ska for critical reading of the manuscript. Numerous discussions with her and Prof. Prof. Tomasz Szoplik, Andrzej Kaim and Andrzej Sikorski are also gratefully acknowledged.

\end{document}